%% file: main.tex
\documentclass[a4paper,preprint,sort, compress]{elsarticle}
\usepackage{graphicx}
\usepackage{amssymb}
\usepackage{amsmath}
\usepackage{lscape}
\usepackage{fixltx2e}
\usepackage{dblfloatfix}

\newcommand{\Ldist}[1]{\mathcal{D}(#1)}
\newcommand{\LdistSpec}[2]{\mathcal{D}_{#1}(#2)}

\begin{document}
\begin{frontmatter}

\title{Transition to congestion in communication/computation networks: from ideal to realistic resource allocation via Montecarlo simulations}

\author{Marco Cogoni}
\ead{marco.cogoni@crs4.it}
\author{Giovanni Busonera, Paolo Anedda, Gianluigi Zanetti}
\address{CRS4, Center for Advanced Studies, Research and Development in Sardinia, Edificio 1\\ 
Parco Scientifico e Tecnologico della Sardegna, 09010 PULA (CA - Italy)}

\begin{abstract}
We generalize previous studies on critical phenomena in communication networks by adding computational capabilities to the nodes to better describe real-world situations such as cloud computing. A set of tasks with random origin and destination with a multi-tier computational structure is distributed on a network modeled as a graph. The execution time (or latency) of each task is statically computed and the sum is used as the energy in a Montecarlo simulation in which the temperature parameter controls the resource allocation optimality. We study the transition to congestion by varying temperature and system load.
A method to approximately recover the time-evolution of the system by interpolating the latency probability distributions is presented. This allows us to study the standard transition to the congested phase by varying the task production rate.
We are able to reproduce the main known results on network congestion and to gain a deeper insight over the maximum theoretical performance of a system and its sensitivity to routing and load balancing errors.
\end{abstract}
\end{frontmatter}

\input{intro}
\input{method}

\input{results}

\input{conclusions}
\clearpage
\bibliographystyle{IEEEtran}
\bibliography{IEEEabrv,DISMOD2}
\smallskip

\end{document}

%% file: intro.tex
\section{Introduction}

The research field of critical phenomena in random networks, originally confined
to theoretical physics and mathematics, has now increasingly common applications
to computer science, biology, traffic engineering and social sciences
\cite{Dorogovtsev,Barabasi2000,Barabasi2009}.
In this work we address the problem of congestion in computer networks.
However, the same approach can be easily applied to traffic flow, container
logistics and energy distribution \cite{Dorogovtsev}.

Statistical mechanics is often used to understand both the topological growth
properties of networks \cite{Bianconi}, and their dynamical behavior
\cite{Molera2008}. Particularly interesting is the phenomenon of congestion
\cite{Arrowsmith2004}.
It has been observed, both in real networks \cite{Barabasi2009,Watts1998} and in
model communication networks \cite{Zhao2005}, that systems only behave
efficiently when the amount of information handled is small enough with respect
to the combined peak capacity. The network collapses above a certain threshold
and part of the information is accumulated and remains undelivered over large time
periods, or it is simply lost \cite{Arrowsmith2004}.

Ohira and Sawatari \cite{Ohira1998} first studied the onset of a phase
transition within a simple model of communication network under different rates
of traffic generation. Their model was defined by a set of nodes on a two
dimensional lattice and controlled by a set of parameters such as packet
generation rate and routing rules. The expected phase transition was observed by
means of the average time interval between packet arrivals at their destination
during a dynamical simulation over time. In 2001, Sol\`e and Valverde
\cite{Valverde2001} slightly extended the model and studied the time-series
dynamics of the number of packets at individual routers, and found the behavior
governed by a set of power laws when approaching the critical point.
In 2006, Chen and Wang explored the effects of network structure and routing
strategy on network capacity \cite{PhysRevE.73.036107} and Gong et al. in 2008
analyzed what happens to network capacity when the available resources are
optimally allocated for Erd\H{o}s-R\`enyi and Barab\'asi-Albert (BA) network
topologies \cite{Gong-Lai-optimal}.

The model presented here tries to build on previous work, where only the network
communication was considered, by: (i) seamlessly including the computation time
(information processing in general) into the total latency; and (ii) avoiding
the explicit temporal evolution by considering a (mean field) average traffic
over a given instant.
By doing so, the typical problem of the queue size of
dynamic simulations is entirely avoided \cite{Arrowsmith2004a} and the focus is
set on the average behavior of the system.

The model can be used to estimate the finite-difference latency derivative for
each computational node and network link and thus to catch possible local
overloads that, in the assumption of a steady state traffic, indicate a
potential long term catastrophe, since some component of the network are not
able to finish their workload within the specified time window.

The explicit inclusion of the computation in the model is expected to be crucial
to the description of real-world situations, for intance applications where a
complex task can be executed over multiple geographic facilities.
An advantage of binding networking and computation is the ability
to manage them concurrently: the optimization of network routing could in
principle be detrimental to the ideal execution time of the associated
processing activity. A multi-objective approach would be needed when such
competing requirements exist. In this paper we present a model that tackles the
optimization problem as a whole, thereby obtaining the best possible total
execution time for the task set from request to delivery.

The model allows an extremely fast evaluation of the latency unfolding the possibility of finding the network's optimal resource allocation. The optimization is performed after setting the network topology and the random traffic distribution \cite{Danila2009,Arenas2008}. 
Moreover, by raising the final temperature in the Metropolis Montecarlo (MC), we estimate the sensitivity of the system performance with respect to the ``amount of non-optimality'' in the allocation of resources (routing and processing combined).
After showing that our approximate model correctly predicts the qualitative behavior observed by means of dynamical simulations, we propose to adopt the ground-state solution found by the Simulated Annealing (SA) as a baseline for comparing any real-world routing algorithm: a performance metric could be defined inversely proportional to the temperature at which the Montecarlo should be run to obtain the same level of efficiency.

We present the behavior of the global latency and probability distribution of execution times with respect to system load and resource optimization for two different topologies: a two-dimensional lattice and a BA random network. We observe a transition to the congested phase when crossing a Pareto curve which is a function of system load and temperature.  

%% file: method.tex
\section{The Impulsive Load Network Model}
The model considers three basic components: (i) the physical support for the
communication/computation process; (ii) the discrete tasks that are
transported/executed; and (iii) the limited capability of the links/nodes to
handle each subtask.
The communication network is mapped onto a graph where nodes mimic the
communicating agents (i.e., routers and servers in a computer network) and the
links between them represent communication lines.  Each node is characterized by
a specific computational capability (which can be zero: i.e., a pure router) and
each link has its own nominal latency and bandwidth. The computational
capability is measured by the number of executable task requests allowable per
time unit.

With respect to the network, each link transports the information at nominal speed when its bandwidth is larger than the total requests of the tasks running on it (which is equivalent to the betweenness centrality measure of the link).
Both for networking and computation, whenever the load surpasses the available resources, a linear slowdown/overload factor is introduced to take into account the increased transport/execution time of each subtask.

Each link is modeled to be able to support several concurrent communications without overloading, while each computational node can execute more than one subtask at the same time.

Within the impulsive model, one computes the total time (latency) needed to
carry out all the system load that is assigned at the beginning. In this way,
the response of the system to an impulse of workload is evaluated at optimal (or sub-optimal) load-balancing and routing. Sub-optimality is obtained by increasing the value of a simulation parameter (the temperature $t$).

\subsection{Network topology}
\begin{figure}[htp]
\begin{center}
\begin{tabular}{ccr}
\includegraphics[width=4.5cm]{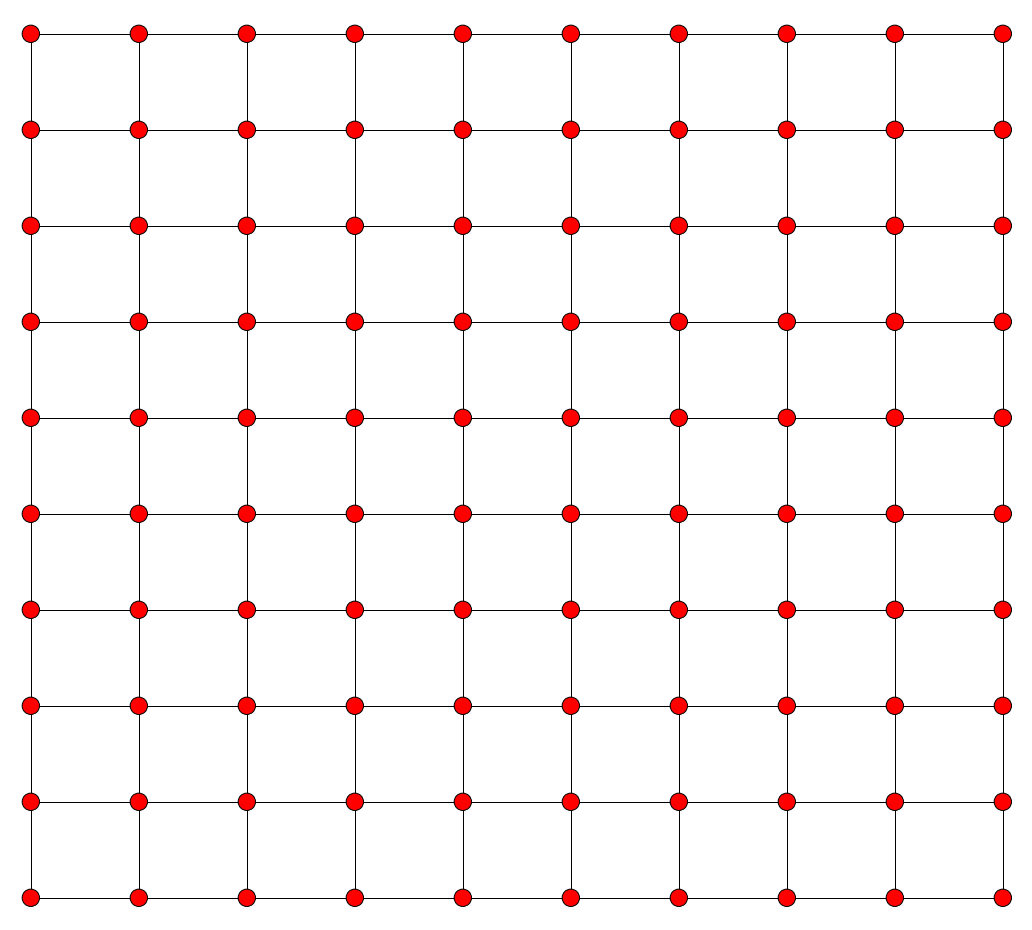} & \includegraphics[width=4.5cm]{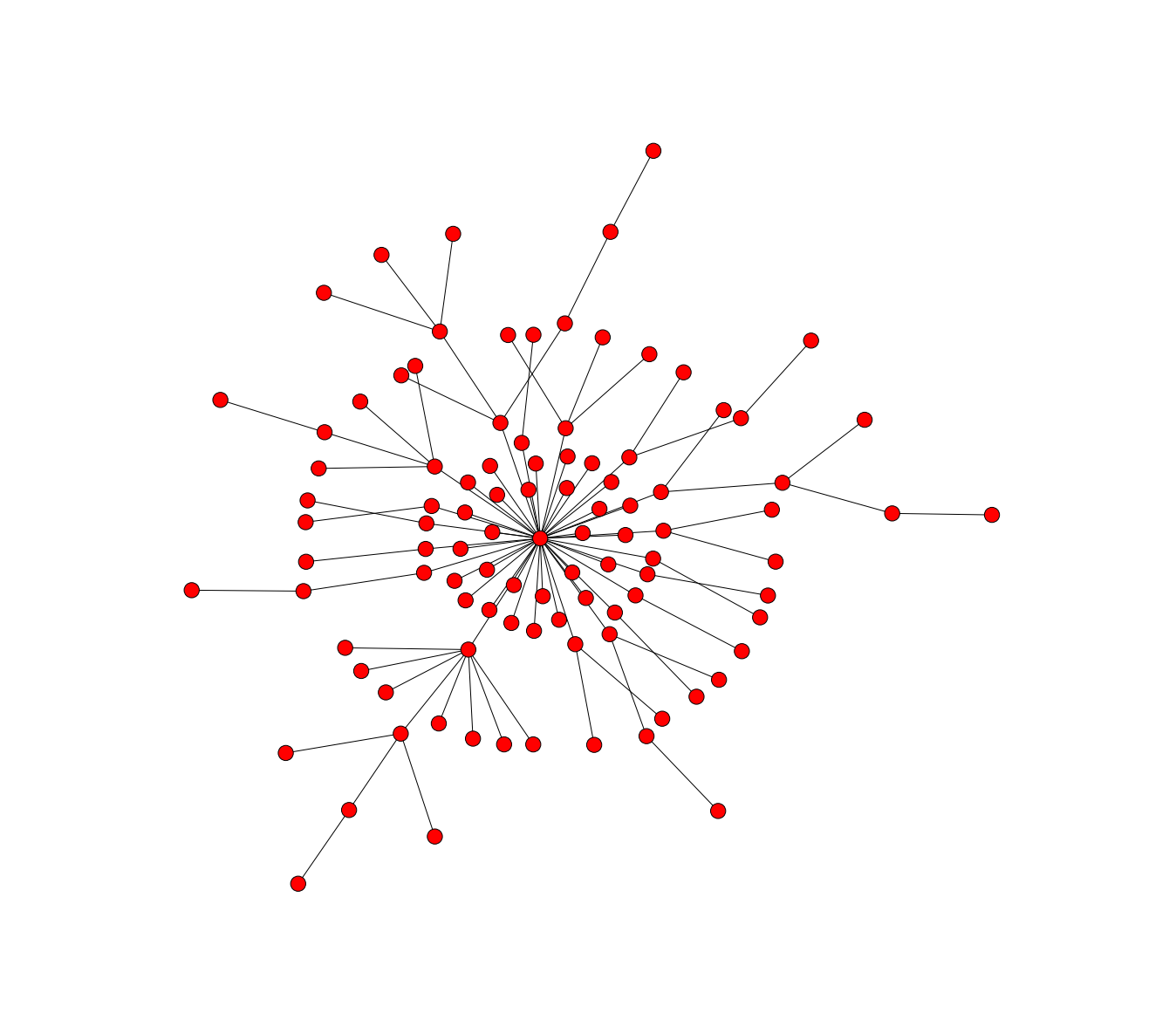} \\
Two-dimensional lattice & Barab\'asi-Albert \\  
\end{tabular} 
\caption{Two-dimensional lattice and Barab\'asi-Albert random network with $M=100$ nodes used for the simulations.}
\label{fig:topol}
\end{center}
\end{figure}

We concentrate on two different topologies depicted in Fig.\ref{fig:topol}:
a simple two-dimensional lattice with no periodic boundary conditions and a Barab\'asi-Albert random network, both with $M$ nodes.
The set of Barab\'asi-Albert random graphs with given connectivity parameters form a statistical ensemble \cite{Bianconi} so several simulations with different disorder realizations are needed to obtain the ensemble average.
The network connections are characterized by: (i) the nominal link latency
$D_{nm}$ between two adjacent nodes $n$ and $m$; and (ii) the attribute $Q_{nm}$
representing the bandwidth capacity. Each node possesses a computational power
$Q_{n}$ representing the amount of computation that can  be executed in
a time unit. In general, one-index parameters (i.e., $Q_{n}$) are associated
with node attributes, whereas two-index ones (i.e., $D_{nm}$ and $Q_{nm}$) refer
to links.

\subsection{Traffic properties}
The total load, described by a set $\{T^{k}\}_{k=1...N}$ of $N$ tasks, can be generated by following any probability distribution for both the computational workload and for the number of subtasks composing each task. Each task is assigned a starting node followed by a sequence of computational stages which it should complete before delivering the result to the final node.
In this paper we adopt the following task structure for every simulation: \begin{equation}
T^{k}=\{n_0^k \xrightarrow{p^k_0} n_1^k\xrightarrow{p^k_1} n_2^k\xrightarrow{p^k_2} n_3^k\}\label{eq:tasks}
\end{equation}
with a total of four stages, the first and the last ones are the input and output nodes while the two in the middle perform the actual processing. Their associated workloads are $$W^{k}=\{w_{a}^{k}, w_{b}^{k}\},$$ ($a$ and $b$ are generic labels for the stages and not node indices) and the $w_{i}^{k}$ factors represent the computational impact of the $k$-th task to the $i$-th node. Each task has a set of workloads impacting the nodes on which it is mapped, but another quantity $w^k_{nm}$is needed to indicate the amount of information transferred along its path from the origin to the end (a constant, unitary, flow of information is assumed for simplicity).

The $p_{i=0,1,2}^k$ over the arrows in Eq.\ref{eq:tasks} describe the
inter-stage paths.
%
To connect each stage with the next, an arbitrary-length, simple path
(without loops) is allowed. Degenerate situations for which any
subset (or all) of the stages coincide on a single node are possible.  The
initial and final nodes of all tasks are spread over the network by following a
uniform distribution. Whenever a task passes through a node without doing actual
processing on it, none of its computational resources are used.

\begin{figure*}[htp]
\centering
\includegraphics[width=12cm]{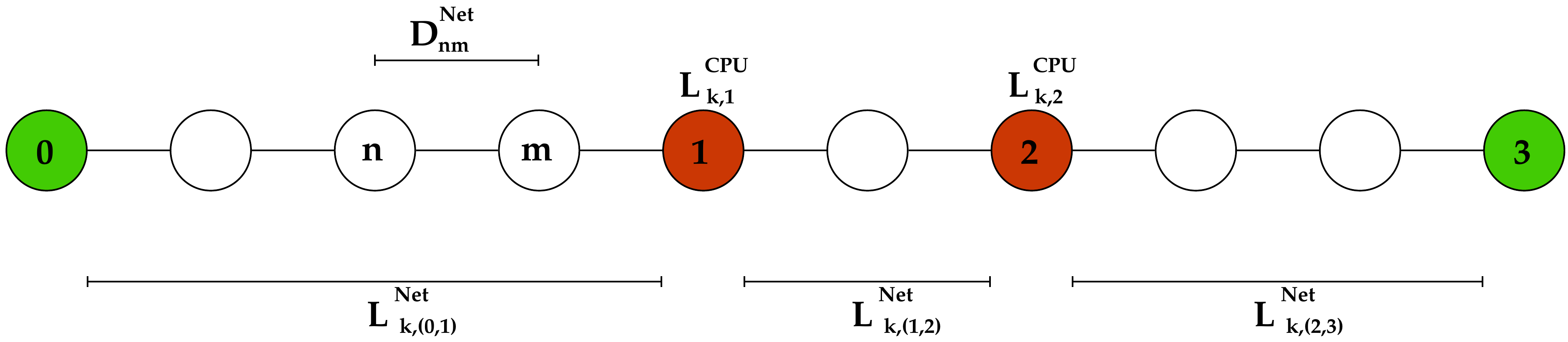}
\caption{Task structure. Green circles are input and output nodes. Red circles represent the processing stages. White circles are nodes used as simple routers between elementary network hops.}
\label{fig:task_graph}
\end{figure*}

\subsubsection{Routing}
From the initial node through each stage and until the final node, the information is passed to the next step via a sequence of network links: the length of each sequence is, in principle, bounded only by the total number of links in the network (no loops are permitted). This structure is shown in Fig.\ref{fig:task_graph}. Routing details are stored in $P^k=\{p_0^k, p_1^k, p_2^k\}$.
Short paths can be slightly preferred to speed up convergence (see the optimization description below for details on the path generation). There are no low level routing rules in the model: the entire space of allowed routes is theoretically explored.
The initial set of routes is chosen either randomly or by computing the weighted shortest paths at the beginning of the simulation: In general this state corresponds to high latency values.

\subsection{Latency computation}
The time needed to complete each task $T^{k}$ (i.e., its latency) is the sum of two components:
$$
L_k^{total} = L^{CPU}_k + L^{net}_k,
$$
\begin{itemize}
\item the network latency for the $k$-th task $L^{net}_k$ is the sum of the delays $L^{net}_{k, h}$ introduced by the paths $p_h^k$ that connect each subtask;
\item the computation latency for the $k$-th task $L^{CPU}_k$ is the sum of the delays $L^{CPU}_{k,n=n_1^k, n_2^k}$ due to all subtasks.
\end{itemize}

\subsubsection{Network latency}
The network latency introduced by a simple direct link $n\Leftrightarrow m$ is the product of a (i) baseline latency term $D_{nm}$, which can be considered as the nominal link latency value between two adjacent nodes $n$ and $m$ and (ii) of a dynamic term dependent on the current network load $S_{nm}$.
$$
L_{k, nm}^{net}=D_{nm} S_{nm},
$$
in which the factor $S_{nm}$ is the ratio between the requests and the maximum capability:
$$
S_{nm}=\frac{\sum_{\forall k}\sum_{h=0,1,2}\sum_{nm\in p_h^k}w^{k}_{nm}}{Q_{nm}},
$$
when the link is overloaded (the sum is over all $k$ tasks using the $n\Leftrightarrow m$ link). Whenever the sum of all requested resources of all tasks using the link between $n$ and $m$ is smaller than its total capability $Q_{nm}$, the ratio is set to $1$.
The $S_{nm}$ factor is proportional to the betweenness centrality measure (sometimes called {\it load}) of the $n\Leftrightarrow m$ link computed with respect to all tasks on the network.
Each element $nm$ in the load matrix $S_{nm}$, when larger than $1$, is used to uniformly slow down {\it all} the subtasks sharing the $nm$ link (the uniformity is due to the lack of time granularity).

In the end, the total latency due to the network transport for the $k$-th task is the sum of the single terms:
$$ L^{net}_k = \sum_{h=0,1,2} L^{net}_{k, h} = \sum_{h=0,1,2} \sum_{nm\in p_h^k} L^{net}_{k, nm}$$

\subsubsection{Computational latency}
The latency contribution due to the computation within each node is modeled in a
very similar way. Each task $T^{k}$ is mapped over a set of computing nodes (at most two for two-stage tasks). The computational slowdown $S_{i}$ for each node $i$ can be computed as the ratio between the sum of the requested resources
$$\sum_{\forall k} w_{i}^{k},$$ ($k$ is the task index) and the computational power $Q_{i}$ of the node:
$$
S_{i}=\frac{\sum_{\forall k}w_{i}^{k}}{Q_{i}}.
$$
Therefore, the time actually spent by each task $k$ to perform the computation will be the sum of its components, each with its penalty $S_i$:
$$
L_{k}^{cpu}=\sum_{s=1,2} S_{n^k_s} w_{n^k_s}^{k} Q^{-1}_{n^k_s},
$$
where the $Q^{-1}_i$ represents the execution time of an elementary task by a free node $i$.

\subsection{Optimization by Simulated Annealing and Montecarlo sampling}
The definition of a performance measure such as the global latency (i.e., the total execution time), which is extensive and can be regarded as the energy in a standard Montecarlo simulation, allows us to explore two important aspects of networks near congestion:
\begin{itemize}
\item finding the configurations with the lowest global latency;
\item using the MC temperature $t$ as a parameter to modify the degree of optimality of the resource management.
\end{itemize}
The first goal is tackled by means of a simulated annealing approach with a slowly decaying temperature which eventually leads to a basin of attraction of configurations with near-optimum performance of the network. The ground state is in general not unique and the SA can in principle be trapped in a higher latency basin. To circumvent this problem we run the optimization several times and then take the best overall solution. Moreover, since the SA behavior is largely determined by the cooling schedule, a compromise between convergence speed and accuracy was found.

Ground states found via the SA algorithm set the maximum performance of a
specific network subject to a steady load: such theoretical performance is
rarely found in reality since computational and routing
strategies have to deal with varying loads and, in general, the full state of
the system is not accessible to the decision agents (see \cite{fastpass} for a similar approach to global optimization applied to a datacenter network). In this work we try to mimic the sub-optimality typical of real-world situations by means of a non-zero temperature within the SA.

\subsubsection{Finding the ground state}
The optimizer tries to find the best possible combined performance for the tasks. After initializing the topology of the network and its resources, a random, uniform, distribution of traffic is generated. The SA procedure starts from a random allocation of the tasks over the network and typically has a poor performance (as in the balls-in-bins statistical problem) \cite{balls-in-bins}.

\begin{figure}[]
\centering
\includegraphics[width=8cm]{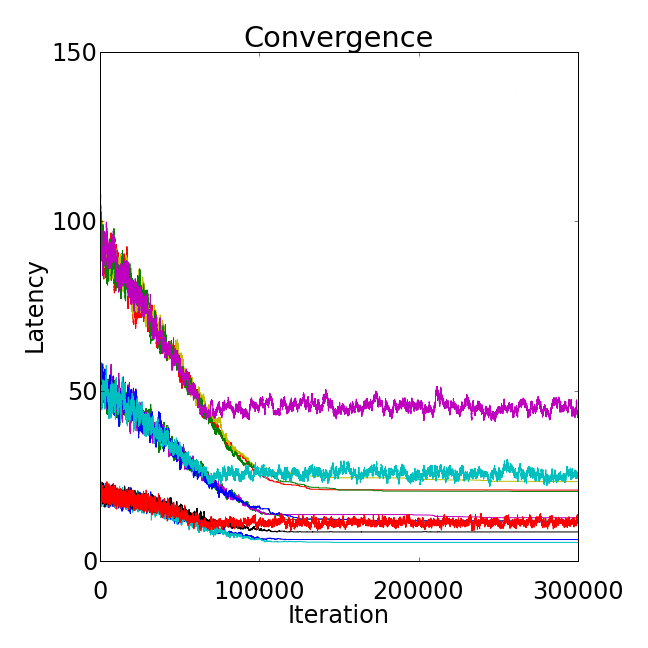}
\caption{Three main families of curves are shown: from top (high task load) to bottom (lowest load) with four different temperatures. The sloping initial part of each run is due to the cooling in the SA. High temperature runs can show higher average latency than lower loads with better resource management.}
\label{fig:convergences}
\end{figure}

The most important part of the SA is the choice of the basic move in
configurational space: In principle it should allow the system to ergodically
explore the states without any bias. Such move would be theoretically
flawless, but the convergence would be very slow when dealing with non-trivial
systems.
The basic move is simple: One of the $N$ tasks is randomly chosen and one of its
subtasks is remapped to one of the $M$ computing nodes. The newly chosen node
will then be reconnected to its ancestor and descendant via a new random
routing. A random choice among the set of all simple paths connecting two nodes
guarantees in principle the exploration of all configurations, but for highly
connected graphs and lattices it would be computationally very expensive. Here
we solve the problem by restricting the pool of possible routes to the set of
all shortest paths augmented by the short paths that avoids a certain number of
nodes belonging to the first set. In this way we can explore an abstract
``spindle'' around each original shortest path with the possibility to modify
its ``thickness'' by changing the number of avoided nodes.
We found that avoiding one or two nodes is enough to efficiently explore the
routing space for the network topologies (and sizes) studied in this work.
As an example of the convergence behavior, in Fig. \ref{fig:convergences}, several convergence profiles over many combinations of temperatures and workloads are shown for the BA network: about $100.000$ SA steps are necessary to approach the ground state.

\subsubsection{Realistic resource management by raising the temperature}
The choice of a standard Gibbs sampling seems natural since it allows to explore a statistical ensemble of states associated to a constant temperature (canonical ensemble). This fact would not be important if our only goal was to find the global minimum of the total latency via a standard SA.
Being able to simulate the system at constant temperature is equivalent, in a
sense, to having a parameter that sets the amount of bad choices in the total
resource allocation. A near zero temperature simulation (after finding the
lowest latency state via the SA exponential cooling schedule) is equivalent to
an ideal allocation. Any small increase in $t$ drives the system to less
efficient states, up to a totally random resource allocation for very high $t$
values.

\subsection{Recovering a coarse-grained time-evolution from impulsive load simulations}
\begin{figure}[]
\centering
\includegraphics[width=12cm]{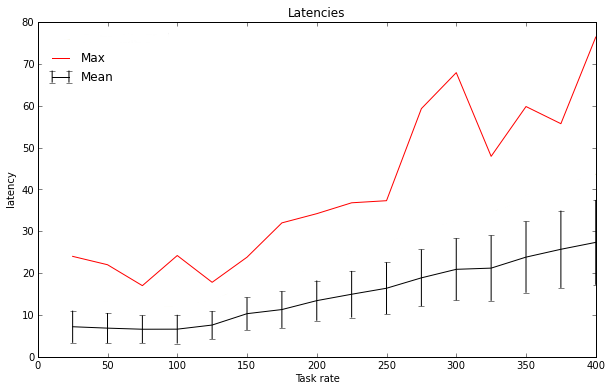}
\caption{Summary of the data presented in Fig.\ref{fig:pdl-vs-load}: Average and maximum values of the latency versus workload.}
\label{fig:pdl-max-mean-vs-load}
\end{figure}

\begin{figure*}[]
\centering
\includegraphics[width=12cm]{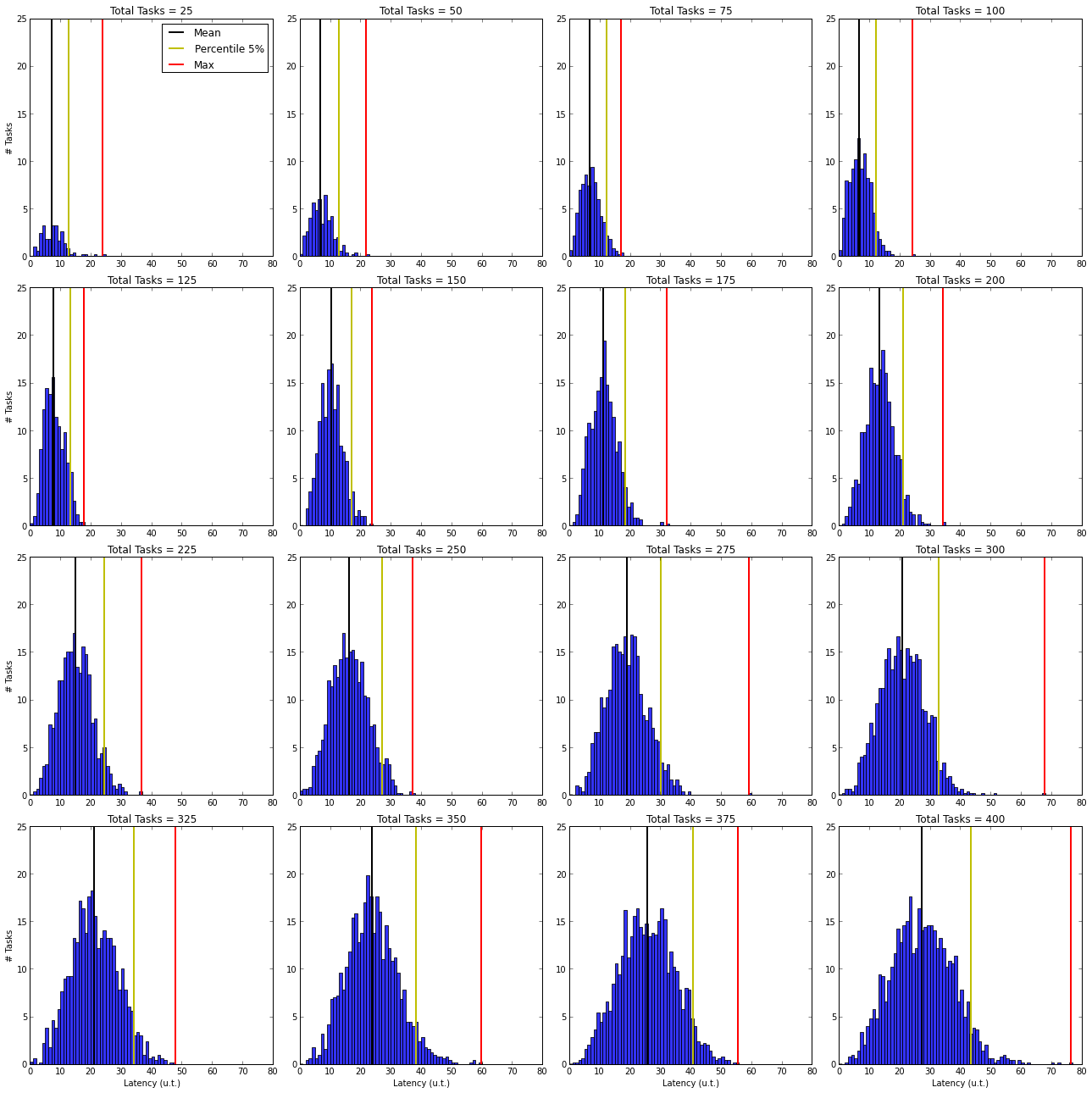}
\caption{Distribution of the occurrences for the latency values (arbitrary time units) and its dependence on the total load on the network (number of concurrent tasks): From the top left (25 tasks) to bottom right (400 tasks). This is the result of a fully optimized 2D lattice network (SA  with zero final temperature) with $100$ nodes.}
\label{fig:pdl-vs-load}
\end{figure*}

In the previous sections we focused on the simulation of the network impulse response, characterized by the latency distribution for a set of specific network loads. The main information obtained from the impulse response is the distribution of latencies $\Ldist{L}$ which, from a coarse-grained standpoint, completely characterizes the behavior of the system. It should be noted that $\Ldist{L}$ is not normalized, and for every value of $L$ it contains the number of tasks to be completed after approximately $L$ time units.
It turns out that the $\Ldist{L}$ approaches a binomial distribution (Fig.\ref{fig:pdl-vs-load}) with its maximum and average values dependent on the total load (see Fig.\ref{fig:pdl-max-mean-vs-load}).
Several simulations for the same load are necessary especially with few tasks: with many tasks, the input/output nodes sample uniformly all possible locations and distances while with few tasks, a single simulation gives a biased specific result.

\begin{figure*}[]
\centering
\includegraphics[width=12cm]{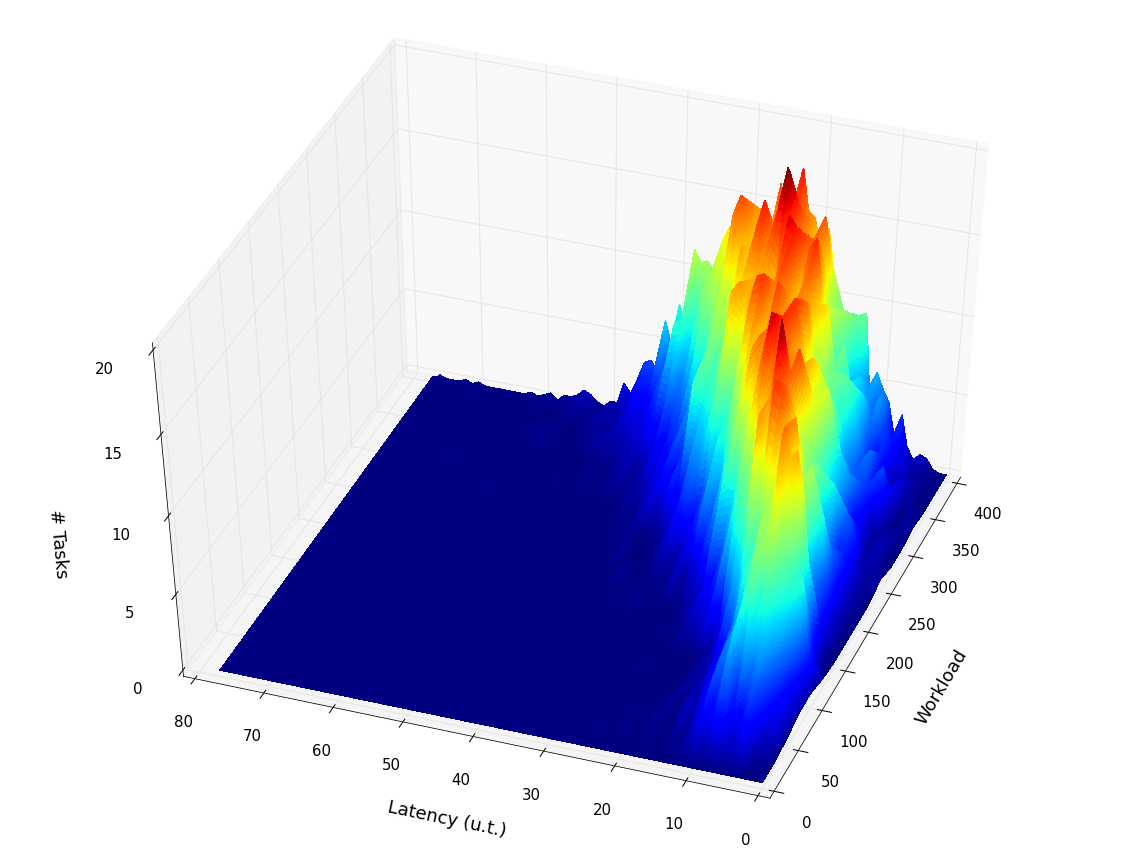}

\bigskip
\includegraphics[width=12cm]{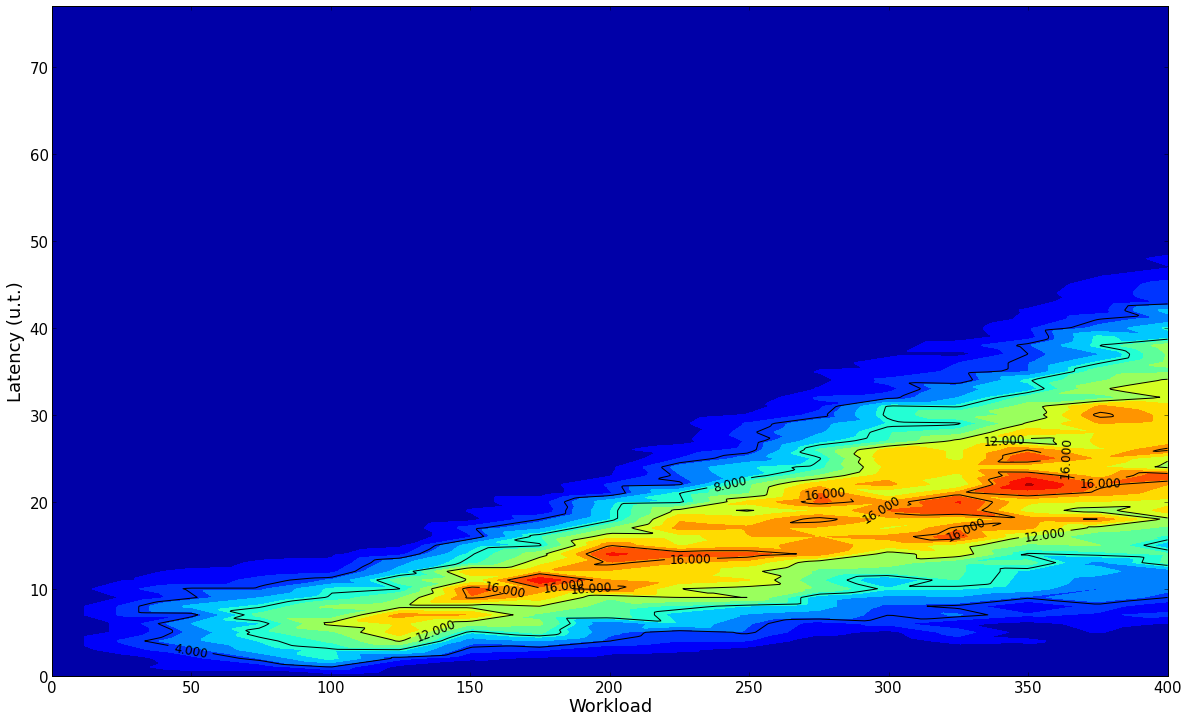}
\caption{Linearly interpolated distribution (from the raw data of Fig.\ref{fig:pdl-vs-load}) of task occurrences for the whole range of latencies and loads. The bottom figure is the contour plot of the top one to better understand the linear-like growth of the average latency and its standard deviation. This data has been obtained at zero temperature, so with ideal resource allocation. In the proposed approximate method to recover the time evolution, the state of the system is entirely defined by the workload coordinate.}
\label{fig:pdl-3D}
\end{figure*}

To recover the latency distribution for an arbitrary load, an interpolation scheme is exploited using the data of Fig.\ref{fig:pdl-vs-load} to get the quasi-continuous distribution shown in Fig.\ref{fig:pdl-3D}. In principle very few impulse simulations could be performed to obtain any $\Ldist{L}$, in practice it is numerically more precise to have several reference points to minimize fitting errors: a two-dimensional linear spline algorithm is used.

A standard network simulation in time is generally packet based
\cite{Ohira1998}. Here, although the tasks play a similar role, the interaction
among them acts in a different way: Packets reside in the queues of the nodes
and they have a specific position in time and space, so they directly interact
only in one point. Our tasks instead, are structured objects occupying a channel
from origin to destination for a time equal to their latency, so the interaction
is among channels and not point-like.

To simulate the coarse-grained time-evolution of any network, we start with no
tasks present at the initial time. During each time unit a new, random, set of
tasks of average size $\lambda M$ is created, where $M$ is the number of nodes
and $\lambda$ is the threshold of a uniform random generator: A task is produced whenever a random number between zero and one is smaller than $\lambda$.  At the same time, a fraction of existing tasks disappears once its destinations are reached. The latter fraction is approximately equal to the sum of completed tasks in each bin $L_i$ of the latency distribution. Specifically, the number of equivalent remaining tasks in the $i-th$ bin after one unit time ($u.t.$) is given by $$\Ldist{L_i}\left({L_i-1u.t.\over L_i}\right)$$ where $${L_i-1u.t.\over  L_i}$$ is the fraction of incomplete tasks with latency $L_i$ after one unit time.
Therefore $$\Ldist{L_i}\left[ {1-{{L_i-1u.t.}\over L_i}}\right]$$ is the number
of equivalent completed tasks for the $i-th$ bin.
So the total amount of completed tasks per unit time is:
$$\sum_{L_i=0}^{L_{max}}\Ldist{L_i}\left[ {1-{{L_i-1u.t.}\over L_i}}\right].$$
This approximate procedure allows us to keep track of the current (coarse-grained) number of tasks:
$$ N_{t+1} = N_t + (\lambda M)_t - \sum_{L_i=0}^{L_{max}}\Ldist{L_i}\left[ {1-{{L_i-1u.t.}\over L_i}}\right]. $$
It is worth noticing that the latency distribution $\Ldist{L}$ depends of course on the current number of tasks $N_t$, so that the correct notation should rather be $\LdistSpec{N_t}{L}$: The distribution at zero temperature is entirely determined by a single scalar (i.e., the instantaneous workload).

%% file: results.tex
\section{Results}
Simulations were run using a two-dimensional lattice (with no PBC) and several instances of BA networks with $100$ nodes and random traffic. Each node is assigned the same unit power.
Tasks are modeled with $2$ processing stages, each with a computational request of one fourth of unit power. This ratio sets the theoretical maximum workload before CPU overload to $200$ tasks. Each point shown on all of the graphs is obtained by averaging over $10$ runs.

\paragraph*{Impulsive Load Simulations}

\begin{figure}[]
\centering
\includegraphics[width=10cm]{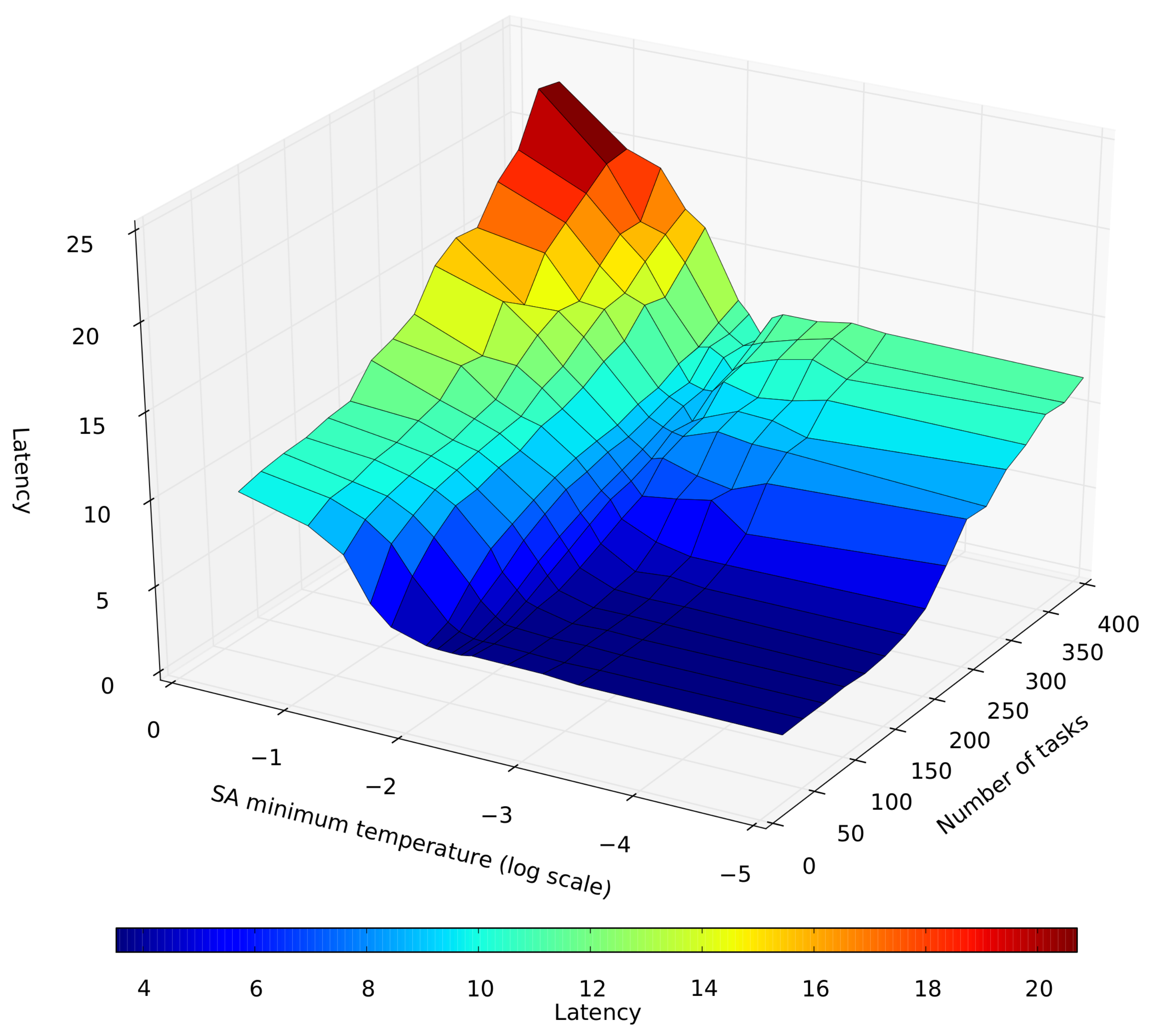}
\caption{Barabasi random network: 3D representation of how the global average latency depends on the traffic and on the Montecarlo temperature: The deep blue basin with low latency can be reached with low traffic and moderately sub-optimal routing or with high load but optimal routing/resource allocation. This system is quite tolerant to sub-optimal choices in the resource allocation.}
\label{fig:3d-phase-diag}
\end{figure}

\begin{figure}[]
\centering
\includegraphics[width=10cm]{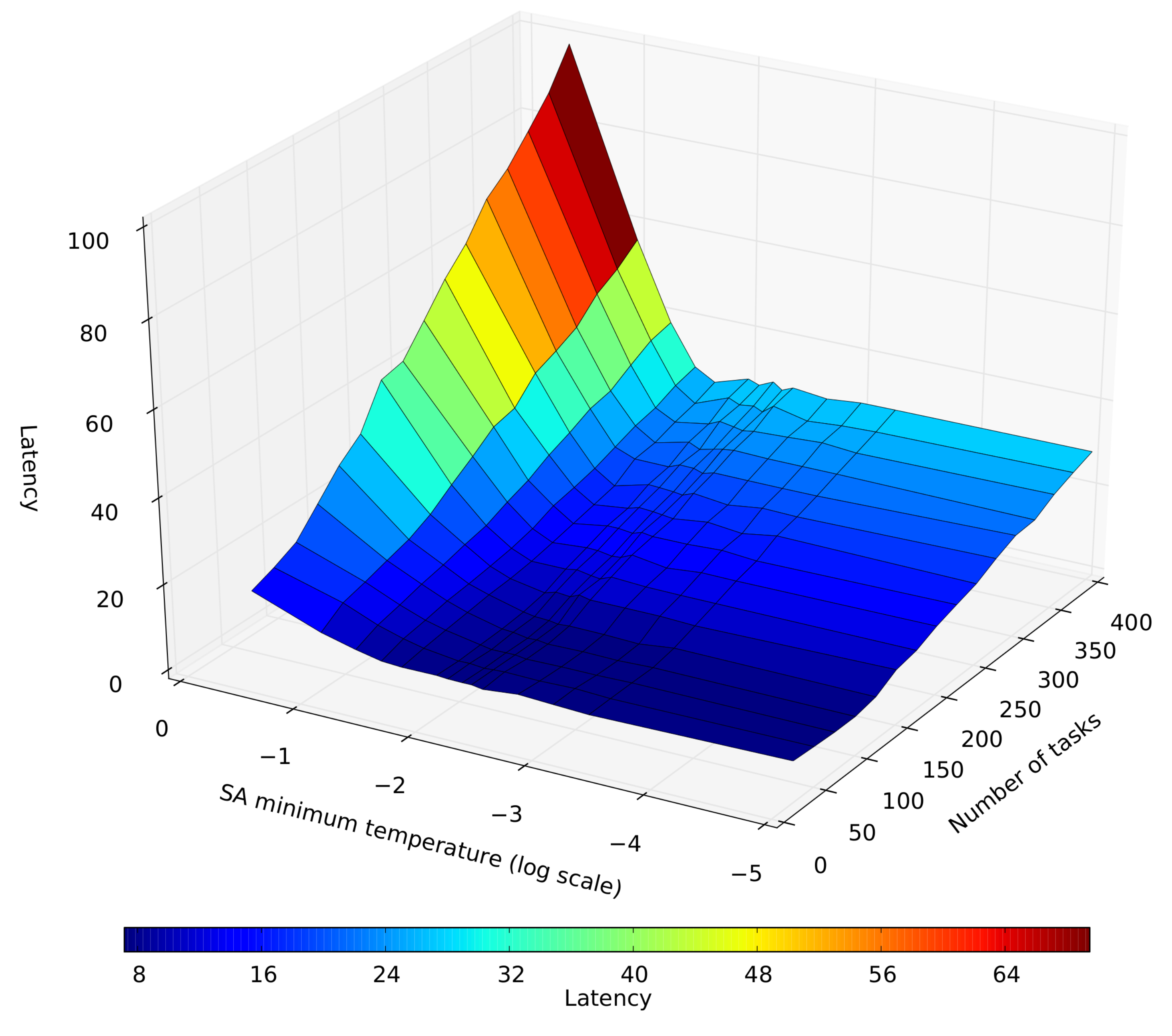}
\caption{Two-dimensional lattice: global average latency dependence on load and temperature for the impulsive load simulations. In this case, the communication network acts as a bottleneck to global performance and high temperatures lead to very latency high values (notice that the latency scale is about 5 times larger than for the BA network).}
\label{fig:3d-phase-diag_mesh2d}
\end{figure}

Both topologies show a marked slowdown well before the theoretical maximum CPU load of $200$ tasks (see Fig.\ref{fig:3d-phase-diag} for the BA network and Fig.\ref{fig:3d-phase-diag_mesh2d} for the lattice), but the global average latency for the lattice is much larger than for the BA, even though both share the same total connectivity: The BA is able to resist higher loads than the lattice since the average path is much shorter, thus the network has a lesser impact on the global latency.


\begin{landscape}
\begin{figure}[h]
\centering
\includegraphics[trim = 50mm 0mm 0mm 0mm, scale=0.3]{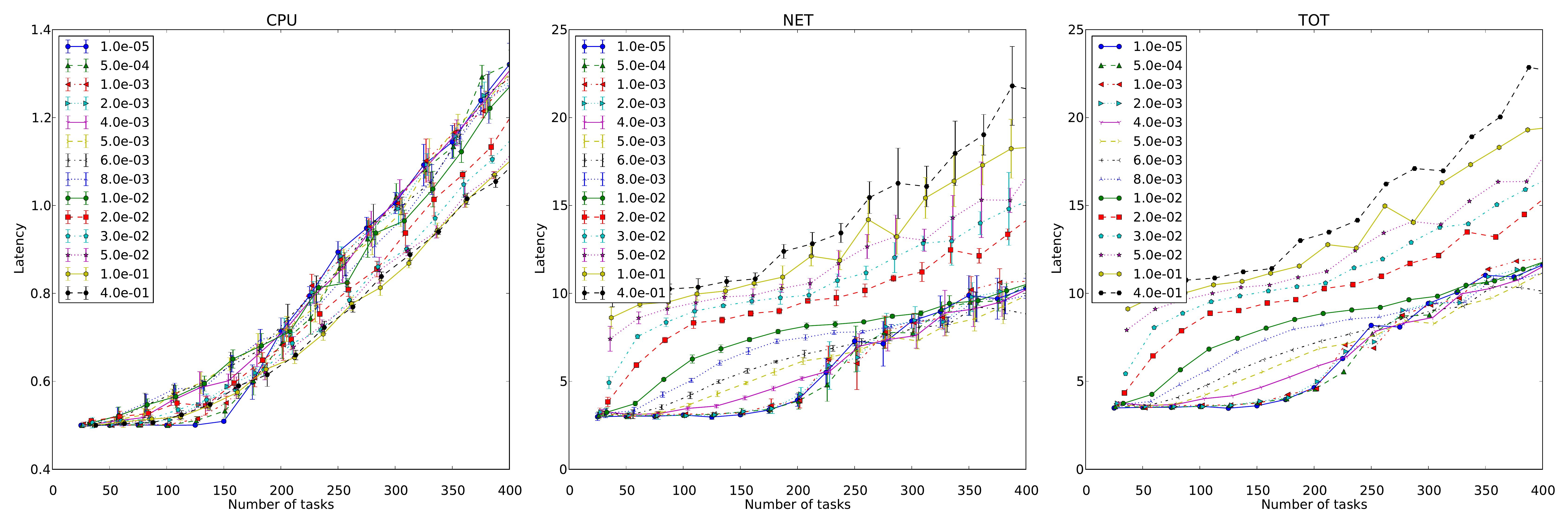}
\includegraphics[trim = 50mm 0mm 0mm 0mm, scale=0.3]{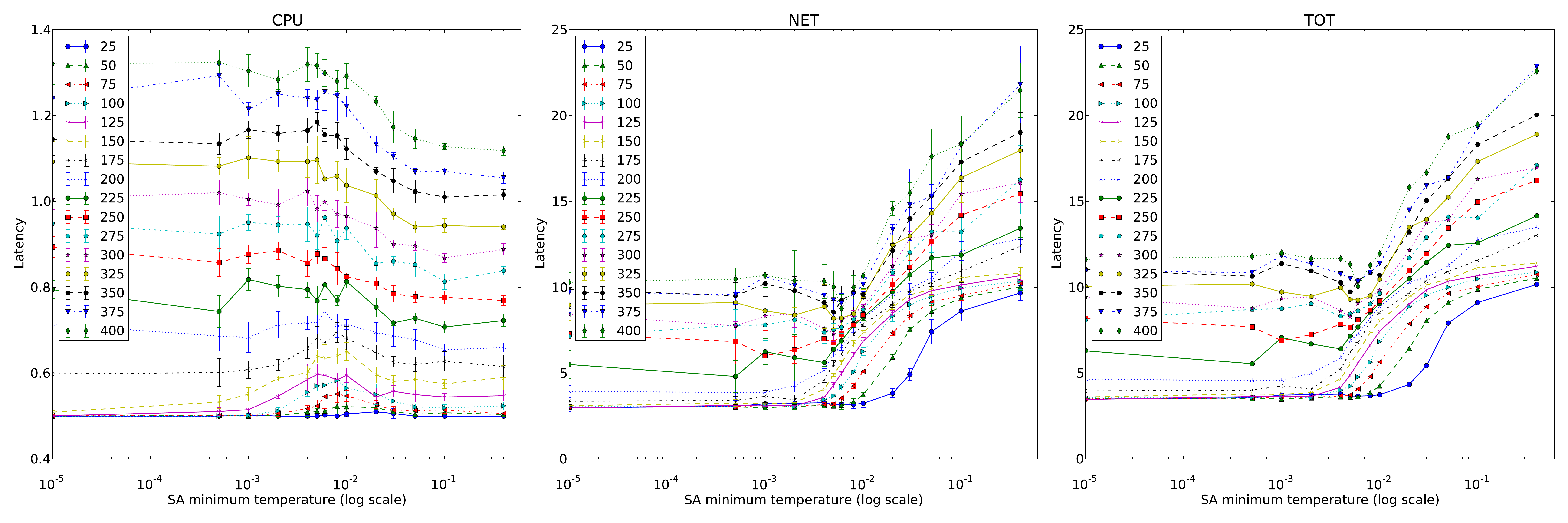}
\caption{Barabasi random network. Top: Total latency (right) and its components (computation on the left and networking in the center box) dependence on the load computed for different Montecarlo temperatures ($10^{-5}, 4\cdot10^{-3}, 10^{-2}, 10^{-1}$). Notice the blue line (lowest temperature) as it gets easily trapped into high latency configurations with a very congested network. Bottom: Total latency (right) and its components (computation on the left and networking in the center box) dependence on the Montecarlo temperature computed for different loads (blue: very light traffic, green: moderate, red: high and cyan: very high).}
\label{fig:2d_phase-diag_barabasi}

\end{figure}
\end{landscape}

\begin{landscape}
\begin{figure}[htp]
\centering
\includegraphics[trim = 50mm 0mm 0mm 0mm, scale=0.3]{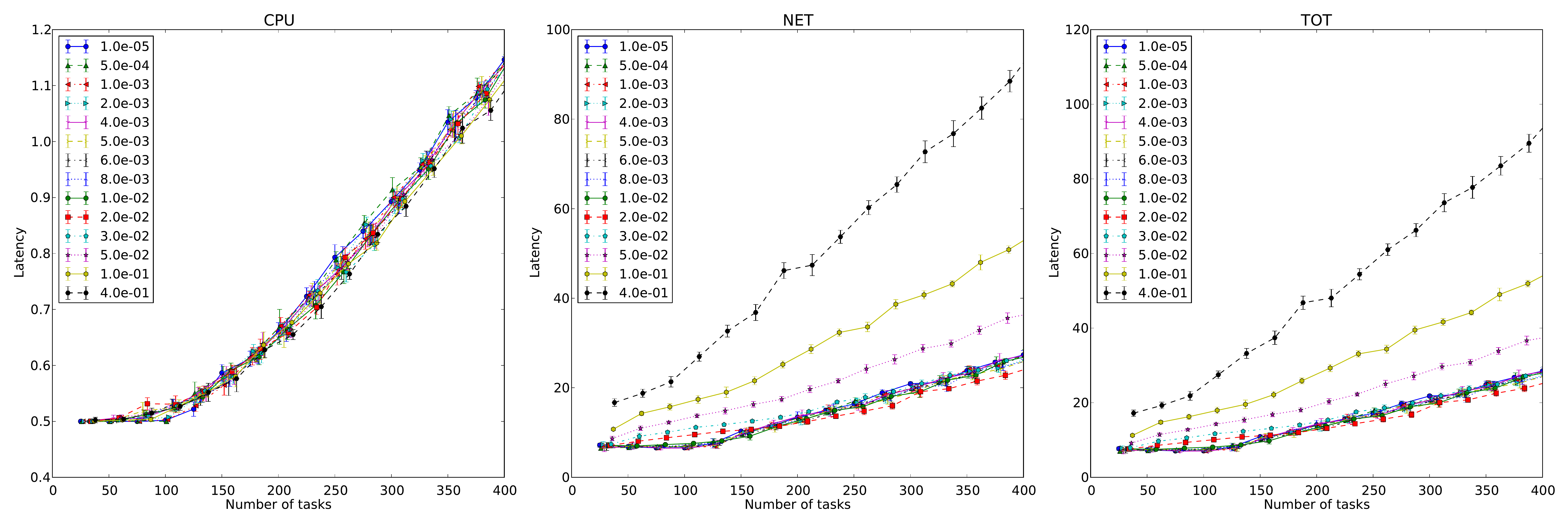}
\includegraphics[trim = 50mm 0mm 0mm 0mm, scale=0.3]{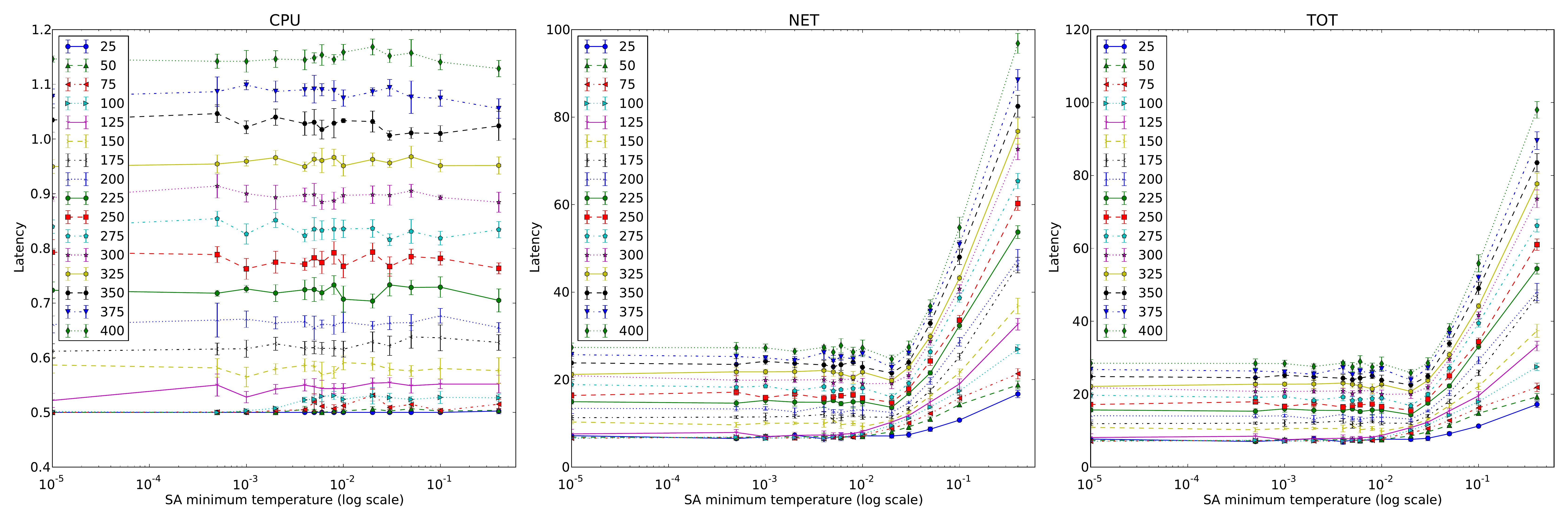}
\caption{2D lattice phase diagram.  Top: Increasing network load: the two diagrams on the left are related to the computation and network latency contribution. The right diagrams shows the total latency. Each diagram has different plots related to different SA minimum temperature - Bottom: Increasing SA minimum temperature: the two diagrams on the left are related to the computation and network latency contribution. The right diagrams shows the total latency. Each diagram has different plots related to different network loads}
\label{fig:2d_phase-diag_mesh2d}
\end{figure}
\end{landscape}

\subparagraph*{Barabasi-Albert}

We first discuss the results relative to the BA network.
As shown in Fig. \ref{fig:2d_phase-diag_barabasi} (top left panel, CPU), the curve labeled with $t=10^{-5}$ (zero $t$) represents the optimal CPU latency at all loads before the transition that sharply occurs above a workload of $N=150$ tasks (note that the maximum theoretical CPU capacity would allow $N=200$ with no overload). Higher workloads lead to a steep linear latency growth with no slope change.
In the central top panel (NET) of Fig.\ref{fig:2d_phase-diag_barabasi}, the curve labeled with $t=10^{-5}$ (zero $t$), shows a smooth transition to slowdown at $N=175$, then the slope increases up to $N=225$ and finally saturates for $N>250$. The saturation implies that the networking latency is comparable to the computational part when ideally optimized.
Raising the temperature, up to $t=2.0\cdot10^{-3}$ nothing changes, then (solid violet curve) the latency worsens for low workloads and grows from about $N=75$: much lower than the theoretical maximum CPU load.
The very high load region $N>250$ does not suffer at all from resource allocation errors up to $t=2.0\cdot10^{-2}$. Another transition is observed for $t\geq2.0\cdot10^{-2}$ (red curve): the whole curve grows (low and high loads) and has a steep slope at $N=25$.
For $t>2.0\cdot10^{-2}$ the latency curve shows an increasing vertical offset, even for extremely light loads ($N=25$).
Fig.\ref{fig:2d_phase-diag_barabasi} (bottom) shows a perpendicular slicing of Fig. \ref{fig:3d-phase-diag} where the behavior of the latency is plot with respect to the minimum temperature of the SA and each curve is associated to a different load.

\subparagraph*{Two-Dimensional Lattice}

Similar diagrams for the two-dimensional lattice topology are depicted in Fig.\ref{fig:2d_phase-diag_mesh2d} (top and bottom).
Lattice simulations show a less rich landscape since network congestion dominates over computational overload for purely geometric reasons: the former is due to both lengthened paths and link slowdown, the latter solely to node slowdown.

One of the main results is that the resource allocation parameter ($t$) has a very strong impact on network performance near the critical load ($N\approx 150$).
The simulated annealing is able to find an extended flat basin with constant latency regardless of workload and temperature. This optimum performance starts to worsen when these two parameters cross a joint Pareto domain wall: The random network is able to withstand much worse resource allocations than the lattice without suffering strong performance degradation.

\paragraph*{Coarse-grained Time Evolution}

Up to now, we concentrated our attention over the impulsive load simulations which characterize how the system reacts to a burst of workload. This kind of information does not directly help to understand the network congestion phenomenon which is inherently something that emerges in a simulation where time is explicitly modeled.

As a showcase of the approximated time evolution approach, we turn our attention to the two-dimensional lattice in which every source node generates a new task whenever a Poisson generator fires at each timestep. Being the network composed by $100$ nodes, the highest task production rate can reach $100$ new tasks per timestep.

\begin{figure}[h!]
\centering
\includegraphics[width=12cm]{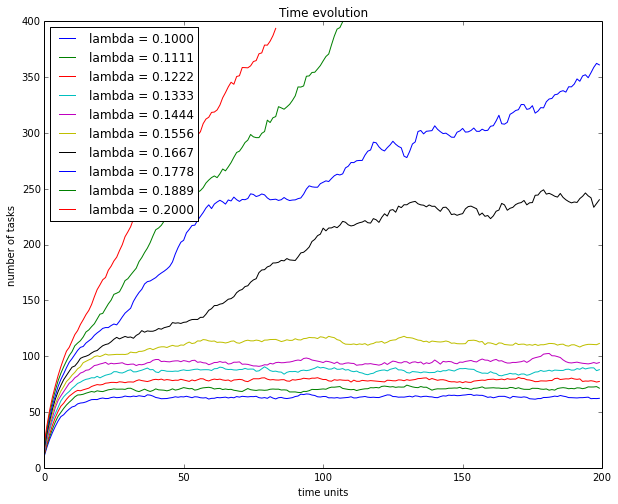}
\caption{Reconstructed time evolution for the two-dimensional lattice with no PBC. A steady number of active tasks is reached after some iterations when $\lambda$ is under the critical value of $\lambda_c \approx 0.16$. For larger $\lambda$ the number of tasks starts to grow linearly with increasing slope.}
\label{fig:time-evol}
\end{figure}

The number of active tasks inside the network is shown in Fig.\ref{fig:time-evol} with respect to time: Several curves for different values of the $\lambda$ parameter (Poisson generation rate at each node) show that for $\lambda > \lambda_c \approx 0.16$, the network undergoes a breakdown passing from a steady state phase to a congested one with monotonic task growth. For low $\lambda$ values the simulation reaches stability after a few tens timesteps after a steep rise when the system is just being filled with new tasks and only few task could have finished. The behavior is qualitatively analogous to classical packet-based simulations and it turns out that the quantitative values are similar \cite{Arrowsmith2004}.

\begin{figure}[h!]
\centering
\includegraphics[width=12cm]{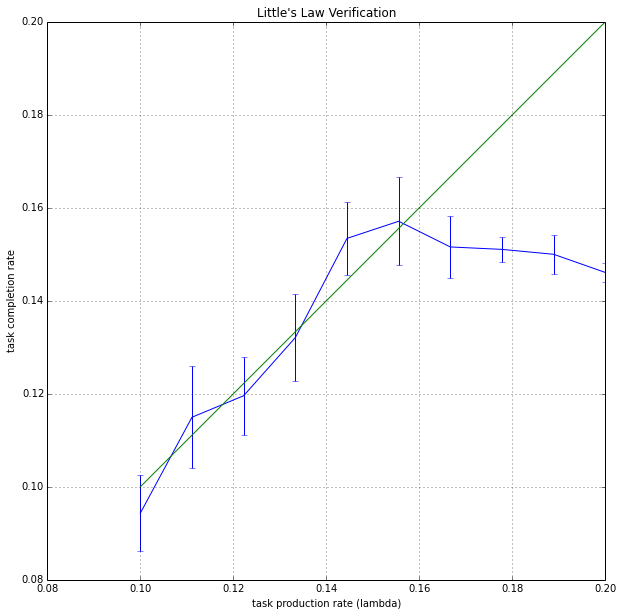}
\caption{Little's law break-up: at high generation rates (congested phase) the network can no longer cope with the production rate since the average latency increases. At $\lambda_c \approx 0.16$ the throughput of the network is maximum, then starts to decrease.}
\label{fig:little-law}
\end{figure}

The same  congestion behavior is well visible when comparing it to a classical Little's Law approximation (the green straight line) in Fig.\ref{fig:little-law}: With a lower than critical generation rate, the network is able to complete a larger number of tasks (y-axis) than the amount of tasks being produced at each timestep (x-axis). Above $\lambda_c$, the network collapses and its ability to complete tasks wanes very quickly as they accumulate faster.
This result, obtained with the novel algorithm, is comparable to the results obtained from dynamical simulations such as those presented in Arrowsmith et al. \cite{Arrowsmith2004}

%% file: conclusions.tex
\section{Discussion and Conclusions}

In this work we present a method to simulate a networked computational system operating under ideal (or realistic) resources allocation. We build upon simplified model networks \cite{Ohira1998,Valverde2001,Arrowsmith2004} by adding computational capability to the nodes. We abandon a detailed time evolution of packets and queues, to adopt a coarse-grained view by initially characterizing the system when subject to impulses of workload. In this framework, the main observable quantity is the probability distribution of latencies $\Ldist{L}$ of a set of computational tasks. We implicitly define an ideal resources allocation as one that minimizes the integral of this probability distribution. This is achieved by applying a simulated annealing approach, whose abstract temperature is used to mimic resources allocation ranging from optimal to sub-optimal, where the former is achieved at zero temperature.

The basic impulsive model is then extended to recover an approximate time evolution of the system by interpolating the execution-time probability distributions over several impulsive workloads. The main idea is that at every timestep we can keep track of how many tasks exist within the system, computed as the difference between newly created tasks and completed ones. This scalar quantity, on average, determines the next $\Ldist{L}$ and so on.

By exploiting this novel technique, we were able to reproduce some fundamental results obtained in the network congestion literature such as the Little's Law breakup \cite{Arrowsmith2004} both for the two-dimensional lattice and the random network.

Even though the present method produces approximate results, it allows to estimate an upper bound for the performance of any network along with an evaluation of its sensitivity to sub-optimal routing or broken links. The temperature parameter could be exploited to define a quantitative reference metric to compare different topologies with respect to maximum performance and resilience.
The same approach could be used to compare the same network when operated with different routing rules.